\documentclass[journal,final]{IEEEtran}
\usepackage{epsfig,rotating,setspace,latexsym,amsmath,epsf,amssymb,bm}

\usepackage{color}
\usepackage{amsmath}
\usepackage{rotating,latexsym,amsmath,amssymb,bm}
\usepackage{cite}

\usepackage{amsthm}
\usepackage{varwidth}
\usepackage{subfigure}
\usepackage{graphicx}

\def\nb0{{\mathbf{0}}}
\def\nb1{{\mathbf{1}}}

\usepackage{bbm}

\usepackage{url}
\usepackage{float}
\usepackage{wrapfig}
\usepackage{dsfont}
\usepackage{wasysym}
\usepackage{hyperref}
\usepackage{balance}
\usepackage[ruled,vlined]{algorithm2e}

\theoremstyle{plain}

\theoremstyle{definition}

\usepackage{thmtools}
\declaretheoremstyle[
  spaceabove=\topsep, spacebelow=\topsep,
  headfont=\normalfont\bfseries,
  notefont=\mdseries, notebraces={(}{)},
  bodyfont=\normalfont,
  postheadspace=1em,
  qed=\qedsymbol
]{mythmstyle}

\declaretheoremstyle[
  spaceabove=\topsep, spacebelow=\topsep,
  headfont=\normalfont\bfseries,
  notefont=\mdseries, notebraces={(}{)},
  bodyfont=\normalfont,
  postheadspace=1em,
  qed=$\diamond$
]{mythmstyle}

\begin{document}

\allowdisplaybreaks

\sloppy

\title{On the Role of 5G and Beyond Sidelink Communication in Multi-Hop Tactical Networks}

\author{Charles E. Thornton, Evan Allen, Evar Jones, Daniel Jakubisin, Fred Templin, and Lingjia Liu

\thanks{C.E. Thornton, E. Jones, and D.J. Jakubisin are with the Virginia Tech National Security Institute (Correspondence: cthorn14@vt.edu). E. Allen and L. Liu are with Wireless $@$ Virginia Tech, Bradley Department of Electrical and Computer Engineering. F. Templin is with the Boeing Company.}}

\maketitle
\thispagestyle{plain}
\pagestyle{plain}
\vspace{-1cm}
\begin{abstract}
This work investigates the potential of 5G and beyond sidelink (SL) communication to support 
multi-hop tactical networks. We first provide a technical and historical overview of 3GPP SL standardization activities, and then consider applications to current problems of interest in tactical networking. We consider a number of multi-hop routing techniques which are expected to be of interest for SL-enabled multi-hop tactical networking and examine open-source tools useful for network emulation. Finally, we discuss relevant research directions which may be of interest for 5G SL-enabled tactical communications, namely the integration of RF sensing and positioning, as well as emerging machine learning tools such as federated and decentralized learning, which may be of great interest for resource allocation and routing problems that arise in tactical applications. We conclude by summarizing recent developments in the 5G SL literature and provide guidelines for future research.
\end{abstract}


\section{Introduction}
Tactical networks are designed to support time-sensitive command and control operations \cite{Rettore2022towards}. These networks must make intelligent use of a variety of communication, sensing, and computational devices and techniques in environments where conventional networking infrastructure is not available and data must be propagated quickly across a large network. The goal of such a network is to achieve an internet-like capability in a hostile operation environment \cite{Burbank2006}.

Since the network topology and operating conditions are expected to change rapidly in tactical applications, the systems deployed at the edge of tactical networks must be robust to changes in the network. Tactical networks are typically based on MANETs, which are designed to promote reliable communication over unreliable radio links. However, due to factors such as the diversity of tactical operations, wide range of equipment utilized, and scale of the network, MANET implementation is necessarily accompanied by many practical challenges \cite{Burbank2006}.

\begin{figure}[t]
    \centering
    \includegraphics[width=\columnwidth]{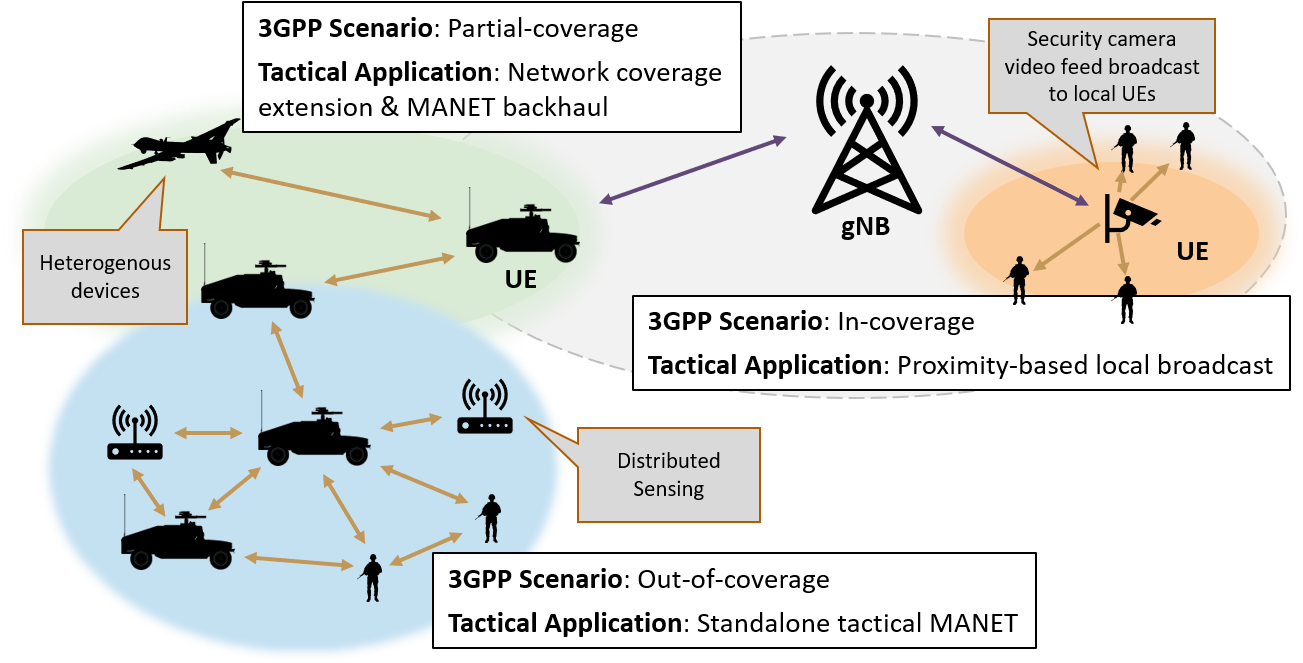}
    \caption{Illustration of tactical applications of 5G Sidelink.}
    \label{fig:scenarios}
\end{figure} 

There is significant interest in leveraging 5G standardized communication capabilities in military networks \cite{elmasry2021operational,bastos2021potential}. Standardized developments in communication technology carry the promise of greatly easing the implementation of tactical networks, which often make use of dated technology. Additionally, 5G technology could greatly enhance the interoperability of military operations, as applications often involve a diverse range of sensing and communication protocols, which could potentially be high-performance but incompatible with one another. However, a limiting factor in the deployment of 5G for military operations is the lack of network infrastructure in battlefield environments. One proposed solution to this deficit is to use specially modified, modestly-sized, 5G base stations for military operation. Another solution, which may be used alongside these small hybrid base-stations, is to use direct device-to-device or “sidelink” (SL) communication, which is currently under vigorous standardization and development. In the present contribution, we explore the viability, potential benefits, and challenges associated with the use of 5G SL in multi-hop tactical networks. Fig.~\ref{fig:scenarios} illustrates tactical applications of SL communications, extending beyond the reach of 5G infrastructure. Ultimately, we envision 5G and beyond sidelink playing a role in a tactical edge enabling land, air, sea, space integration.

5G SL offers the possibility of high data rate communication in mission critical applications in which access to base stations may be severely limited or nonexistent. Some of the benefits of SL are improved spectrum efficiency, flexible radio resource allocation, and compatibility over wide range of available spectrum. While SL communication is not unique to the 5G standard, there are several compelling advantages that 5G SL offers over previous deployments.  For example, 5G NR SL offers a flexible numerology~\cite{campolo2019impact}, wider array of modulation and coding schemes, more flexible HARQ schemes, and base support for spatial multiplexing, allowing for significantly improved link capacity when compared to 4G LTE SL.

 However, we note that several key research and development challenges must be addressed to facilitate successful implementation of 5G NR SL into robust tactical networks. Foremost, multi-hop relaying must be developed further in 5G SL standardization \cite{Bartoletti2022sidelink}. Additionally, 5G waveforms designed for commercial applications, need to be evaluated for resiliency in tactical operation.

 \subsubsection*{Contributions and Organization} This contribution surveys key developments and open research challenges which are expected to be crucial in the push towards 5G SL integration in tactical networks. The paper is organized as follows. Section \ref{se:slbg}, covers relevant history of SL standardization and key technical background. Section \ref{se:multihop} considers challenges in multi-hop routing in SL-enabled networks. Section \ref{se:emulation} surveys recent developments in open-source SL-enabled network emulation. Section \ref{se:integrated} describes potential use-cases and research directions for networks that consider integrated sensing, communication, and computation. Section \ref{se:learning} describes research directions in distributed machine learning relevant for SL-enabled tactical networks. Section \ref{se:concl} provides concluding remarks and surveys open research problems. 

\section{5G SL Preliminaries}
\label{se:slbg} 
\textit{3GPP SL History:} 
D2D technologies have long been in development, being first introduced in 2010 under IEEE standard 802.11p titled \emph{Wireless Access in Vehicular Environments (WAVE)}. An early predecessor to modern sidelink, WAVE technology's contribution was largely limited to dedicated short-range communication for vehicular safety protocols. In 2012, LTE Proximity Services (ProSe) introduced direct-mode communication in 3GPP Release 12. Allowing users to bypass the cellular network infrastructure with the aim of enabling proximity based services and a variety of public safety applications. Additionally around this time Long-Term Evolution Vehicle-to-Everything (LTE-V2X) technology was developed as part of Release 14, expanding upon many of the ideas introduced in 802.11p and adding new functionality to LTE ProSe. In Release 16, 5G-NR V2X SL protocols are standardized. Support for flexible numerologies was added, operating on various spectrum band such as the intelligent transport system (ITS) dedicated band and the licensed band of frequency range 1 (FR1) and FR2. For SL synchronization, GNSS, gNB/eNB and the NR SL UE can be used as a synchronization reference source of a UE. 5G is additionally able to use specific transmit configurations to support Ultra-Reliable Low Latency Communication (URLLC) ideal for important time-sensitive messages.

\textit{5G-NR Channels:} 
NR V2X protocols utilize the following physical channels. The \textit{Physical SL Broadcast Channel (PSBCH)} and its demodulation reference signal (DMRS), the \textit{Physical SL Control Channel (PSCCH)} and its DMRS. \textit{Physical SL shared channel (PSSCH)} and its DMRS. \textit{Physical SL feedback channel (PSFCH)}. \textit{SL primary and secondary synchronization signals} (S-PSS and S-SSS), the \textit{Phase-tracking reference signal} (PT-RS) in FR2, and the \textit{Channel state information reference signal} (CSI-RS).

As outlined in Release 16, SL control information (SCI) is transmitted in two stages. In \textit{stage one}, SCI is carried on the PSCCH containing information which enables sensing operations in addition to the resource allocation information of the PSSCH. 
In \textit{stage two}, the PSSCH serves as the stage 2 SCI carrier as well as serving as the the SL shared channel (SL-SCH) transport channel. Second stage SCI information enables decoding of the associated SL-SCH in addition to hybrid automatic repeat request (HARQ) control information, CSI feedback triggers and other associated support data. 

\begin{figure}[t]
    \centering
    \includegraphics[scale=.225]{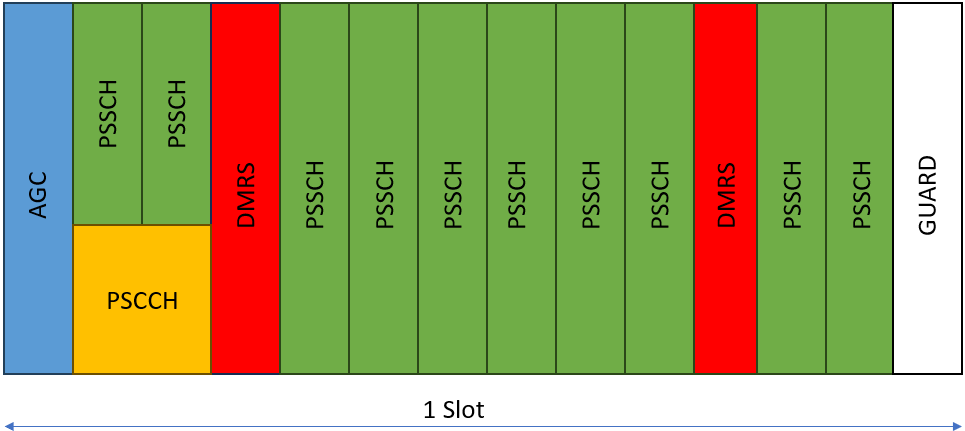}
    \caption{Example slot format of 2-symbol PSCCH, 2-symbol PSSCH-DMRS, and no PSFCH.}
    \label{fig:5G SlotConfig1}
\end{figure}

PSCCH and PSSCH are multiplexed in time and frequency within a slot for short latency and high reliability. DRMS is frequency multiplexed with PSCCH or PSSCH in the corresponding DMRS symbols. Optionally PSFCH can be transmitted at the end of the synchronization slot to enable SL HARQ feedback for unicast and groupcast operations. An example of a supported slot configuration is shown in Figure \ref{fig:5G SlotConfig1}.

\begin{figure}[h].
    \centering
    \includegraphics[scale=.225]{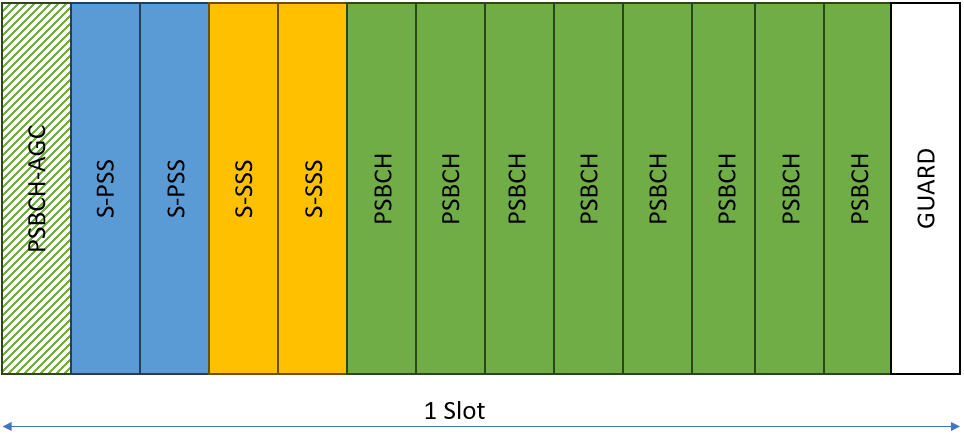}
    \caption{Example slot format of SL Synchronization Signal Block (S-SSB).}
    \label{fig:5G S-SSB}
\end{figure}

\textit{5G-NR Synchronization and Broadcast channel:} 
A slot that transmits the S-SS/PSBCH block is known as a SL Synchronization Signal Block (S-SSB). It consists of PSBCH, SPSS and S-SSS symbols. The period of the S-SSB transmission is 16 frames, and within each period, the number of S-SSB blocks is configurable at the Radio Resource Controller(RRC). The PSBCH symbol then serve as the primary UE data carrier.  An example configuration of a S-SSB block is shown in Figure \ref{fig:5G S-SSB}



\begin{table*}[t]
\caption{5G Flexible Numerology Table.}
\centering
\footnotesize
\begin{tabular}{|c|c|c|c|c|c|c|c|}
\hline
$\mu$ & SCS [kHz] & $N^{Frame,\mu}_{Slot}$ & $N^{SubFrame,\mu}_{Slot}$  & $\tau_{Slot,\mu}[ms]$ & Cyclic Prefix & $N^{Slot}_{Symbol}$ & $N^{SubFrame,\mu}_{Symbol}$  \\
\hline
0  & 15 & 10 & 1 & 1 &Normal & 14 & 14    \\
\hline
1  & 30 & 20 & 2 & .5 &Normal & 14 & 28   \\
\hline
2  & 60 & 40 & 4 & .25 &Normal & 14 & 56    \\
\hline
2  & 60 & 40 & 4 & .25 &Extended & 12 & 48    \\
\hline
3  & 120 & 80 & 8 & .125 &Normal & 14 & 112  \\
\hline
\end{tabular}
\label{tab:5GNume}
\end{table*}

\textit{Resource Allocation:} 5G-NR supports two resource allocation modes. \textit{Mode 1:} Resource Allocation by gNB.
\textit{Mode 2:} UE autonomous resource selection. In mode 1 the gNB schedules to the UE the dynamic grant resources via downlink control information(DCI), or the configured grant resource type 1 and type 2 by radio resource control (RRC) signalling and DCI respectively. In mode 2 the UE performs a sensing operation to determine transmission resources by UE. This is comprised of 3 phases. In \textit{phase 1,} the UE senses withing an given sensing window. In \textit{phase 2,} the UE excludes resources reserved by other UE's. In \textit{phase 3,} the UE selects the final resources within a selection window. 

In Mode 2,  before transmitting in the reserved resource a sensing UE will re-evalaute the set of reserved resources to check if its intended transmission is still suitable. Preemption is also introduced, such that a UE selects new resources even after it announces the resource reservation when it observes resource collision with a higher priority transmission from another UE. 



\textit{Flexible Numerology:} 
A new feature in 5G NR SL is the increased number of available configurable numerologies. LTE and 5G both hold several mutually defined standards such as the duration of a frame and subframe at 10ms and 1ms respectively, and that one Resource Block is 12 consecutive subcarriers. 5G offers additional functionality in the ability to adjust the duration time of one 14 symbol slot.

As seen in Table \ref{tab:5GNume}, adjusting $\mu$  varies the subcarrier spacing according to the following expression $\Delta f = 2^{\mu} \cdot 15[kHz]$. Thus, by adjusting $\mu$ we can adjust the sub carrier bandwidth, which allows for shorter transmission time and more slots per frame. In Table \ref{tab:5GNume}, it is shown that one slot may vary from one slot per subframe up to eight slots per subframe.

\textit{HARQ:} In LTE ProSe, there was no consideration for Hybrid Automatic Repeat Request(HARQ) feedback functionality though limited support was later included though only for blind transmission. Building upon this 5G SL shows a considerable increase in capacity in specific scenarios in part due to its configurable support for HARQ feedback. While LTE then only supported a set 4 blind HARQ transmissions 5G supports variable re-transmission rates from 1 to 32 transmissions. The ability to feedback the needed structure then allows for increased flexibility in the data rates and reliability possible by 5G SL.

\begin{table}[h!]
\centering
\caption{3GPP SL Feature Releases.}
\footnotesize
\begin{tabular}{|c|c|}
\hline
Feature & Introduction  \\
\hline
V2X Introduction  & LTE ProSe  \\
\hline
D2D Introduction & LTE ProSe \& 3GPP R16  \\
\hline
D2D HARQ(Fixed) & LTE ProSe \  \\
\hline
D2D HARQ(Configurable) & 3GPP R16  \  \\
\hline
5G-SL Protocol stack  & 3GPP R16 \\
\hline
Flexible Numerology  & 3GPP R16   \\
\hline
SL Beamforming & 3GPP R16 \& R17   \\
\hline
SL Network Scheduling  & 3GPP R16 \& R17   \\
\hline
SL Relaying  & 3GPP R17 \& R18   \\
\hline
SL Positioning  & (ongoing) 3GPP R18   \\
\hline
SL MultiHop Routing & (Future) 3GPP R19   \\
\hline
\end{tabular}
\label{tab:5GFeatures}
\end{table}

Thus, 5G SL offers a variety of improvements over LTE SL, particularly in configurability. These developments are expected to result in higher capacity operation for tactical communication in harsh environments.  


\section{Multi-Hop Routing in Tactical Networks}
\label{se:multihop}
A ubiquitous challenge in wireless systems is the rapid envelope decay of a transmitted signal. In some cases, boosting transmission power or adjusting the frequency band may mitigate losses, however this often infeasible in tactical systems. One option to then improve data throughput without a change of frequency band nor transmission power is Multi-Hop Routing. In a Multi-hop Network intermediate UE devices can server as relays to the destination UE in cases of environmental challenges. In UE rich environments, a route of relay UE's can be decided which balance interference to other active UE's while improving data rate and throughput to fringe elements \cite{Draper2011cooperative}. Routing decisions are based upon a variety of available link metrics and system limitations such as the maximum number of hops, the transmitted power, path reliability and mutual interference. 

In tactical MANET's, many devices will enter and exit the network across time. Additionally, a MANET network may feature a wide variety of hardware platforms at each node forcing the network controller to account for each devices specific transmission power, supported Modulation Coding Scheme (MCS) modes and more. A standard 5G structure which expects a centrally located gNB base station may struggle to best account for an irregular coverage area of such UE diversity. 

At present, 5G SL supports four scenarios of UE routing. \textit{Scenario 1:} Remote UE is Out Of Coverage(OOC) and UE-to-NW Relay is In Coverage(IC). \textit{Scenario 2:} Remote UE is IC and UE-to-NW Relay is IC. \textit{Scenario 3:} Remote UE is a different cells coverage than UE-to-NW Relay.\textit{Scenario 4:} Two UE's both OOC. 5G SL does not currently support any form of standardized multi-hop routing. As such to implement any of the systems following will require an increase in 5G standardization. This expansion is currently under preliminary discussion as a possible addition in 3GPP Release 19.

\textit{Open Link State Routing (OLSR)} is a proactive link state routing protocol that utilizes hello messages with topology control information statements to maintain network routes. Proactive networks maintain information on all routes concurrently, even when routes are not in demand. The network maintains routing information by instructing each node to periodically broadcast a hello message to all neighboring nodes. This message contains identification information as well as a list of detected neighbors. This allows each node to know 2 hops of information, its neighbors and its neighbors' neighbors. OLSR then instructs each node to select a subset of its neighbors as MultiPoint Relays(MPRs) with the goal of covering all 2 hop neighbors while minimizing redundancy. These MPRs then periodically broadcast Topology Control(TC) messages containing information about themselves and what nodes they cover to aid other nodes in learning about the network topology, in this process non-MPR nodes will selectively rebroadcast TC messages to inform all nodes. Based on these TC messages nodes can then construct a graph of total network topology which they then use to calculate the shortest distance from themselves to any other existing node. 

\textit{Open shortest path first - MANET Designated Router (OSPF-MDR)} is another form of proactive link state routing protocol which is based on one of the earliest link state protocols designed originally for IPv4 wired networks. 
OSPF was later extended to interact with MANET scenarios with the inclusion of the MDR protocol. The Manet Designated Router (MDR) serves to minimize network flooding, they serve this role by allowing network nodes to assign and build a central structure of MDR nodes, these then cut down on Link State Advertisements (LSA) by allowing primary relaying to occur through MDR nodes. Similarly to how OLSR utilizes MPR nodes so that only a subset of nodes must transmit TC information. However OSPF does differ from OLSR in a few ways, namely in the algorithms used to decide the shortest path from each node and in the quantity of of MDR nodes vs MPR nodes in OSPF.

\textit{AODV: Ad Hoc On-Demand Distance Vector (AODV)} is a reactive protocol. Reactive protocols discover routes on an on demand basis. Discovery is performed by first flooding the network with Route Request (RREQ) packets. Each node that receives an RREQ will then maintain information about the source of the RREQ aswell as information on the previous hop link. Once the destination or a node with a route to the destination receives the RREQ a responding route reply (RREP) packet will be sent back to the source following the path the RREQ followed. At the source then all possible paths to the destination will relay a RREP packet in which case the shortest path is selected. From this point this route will then be utilized to send data in a hop-by-hop fashion to the destination. Should the route change and transmission be interrupted a route error (RRER) packet is sent to clear the path. On-Demand routes are maintained only for a fixed time before being expiry and the process restarts. 

As AODV improves network congestion and overhead by only maintaining routes when one is requested, it also incurs larger overhead delay as each route must be discovered prior to transmission. Conversely, OSPF-MDR and OLSR increase network congestion due to maintaining network topology but improves general network awareness and decreases delay in transmissions. 

\section{Open-Source Network Emulation and Prototype Developments}
\label{se:emulation}
A crucial step towards evaluating 5G SL viability for tactical applications is development of scalable and high-fidelity modeling and simulation tools. In this Section, we highlight recent developments in software-based testing relevant for 5G military networks and survey future needs towards full integration of 5G SL in military-grade network emulators.

In general, there are three ways to study the performance of 5G networks. In \emph{link-level} simulations, comprehensive performance analysis is carried out by modeling the individual data links and physical layer components. Of particular interst to the present study, a MATLAB-based open source link-level 5G SL simulator has been developed, which complies with the 3GPP 5G SL standards and offers flexible control over various PHY configurations \cite{liu20225g}. 

However, as the network size grows, the computational resources and time required to model the many-to-many connectivity, as well as complex noise and interference modeling, becomes infeasible. As a more practical alternative, \emph{system-level} simulations may be used to calculate network-wide performance metrics. Finally, \emph{network-level} simulation may take into account various network components such as core network and radio access networks, and focuses on the performance of the entire network. An example of a network simulator is ns-3, which has both LTE and 5G modules available. 

The Common Open Research Emulator (CORE) is an open-source software platform which enables the development of virtual networks. CORE uses the representation of a real computer network by running abstract models in real-time on one or more machines. Live CORE emulations can interface with physical networks and routers. Military-grade and large-scale network emulators, like EMANE and CORE, have been used in several efforts to study the feasibility of battlefield readiness for LTE and 5G deployments \cite{Ryu2022emane,Le2022physical,elkadi2023open}. While both CORE and EMANE provide platforms to integrate various radio and network models, these emulators require as input link performance metrics, stored in the form of look-up tables. Such link performance data is relatively straightforward to obtain for known scenarios in which each node's mobility has a pre-determined trajectory. However, in environments where the wireless channel dynamic can be expected to change drastically, the LUT approach can be rendered ineffective. 

When the wireless environment changes drastically, there are several options. One could execute a full link-level simulation for the new environment. Another is to apply a Physical Layer Abstraction (PLA) method to produce a new LUT that acceptably approximates the new environment. Finally, one could use existing PLA models to project existing characterizations to a new LUT. Given the difficulty associated with performing full link-level simulations for each node in the network every time the radio environment changes, the development of accurate PHY abstraction and projection techniques is crucial for the emulation of SL-enabled tactical networks.

To this end, a few existing works have considered PLA techniques for 5G SL, namely \cite{Le2022physical,cao2023efficient}. However, there is remaining need for SL PLA techniques in tactical networking scenarios, as well as a continual need for development as new SL technologies are implemented. Further, more comprehensive PLA work, considering the impact on additional PHY parameters, such as the number of resource blocks is needed, as noted in \cite{cao2023efficient}.

In \cite{elkadi2023open}, a prototype NR SL physical layer is developed using open-source software, namely OAI, integrated into an USRP B210 SDR platform, and tested over-the-air. The prototype system is focuses on mode 2, the partial coverage scenario, and follows 3GPP Release 16. All physical SL channels and signals are tested, and the OTA radio tests show performance which align well with the simulation results.

\subsection{Internetworking Considerations}
\begin{figure}
    \centering
    \includegraphics[scale=0.75]{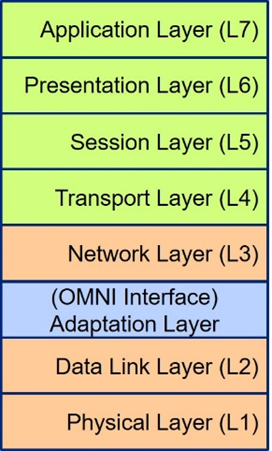}
    \caption{Augmented OSI Reference Model}
    \label{fig:templin}
\end{figure}

5G SL MANETs often comprise mobile nodes that encounter one another within a common tactical operating region with no prior arrangements. In many instances, the MANET is coordinated through the cooperative relaying and forwarding actions of only the equipment carried by the constituent nodes themselves with little or no support from critical infrastructure. The MANET routing function therefore must occur at a layer below standard Internet routing, since there is no notion of a shared Internet subnet maintained by common infrastructure reference points. This means that each MANET node must somehow configure a unique local address for operating the routing protocol that is  assured to avoid conflicts with the addresses used by other nodes.

The AERO/OMNI technologies therefore introduce a new “8th layer” in the 7-layer OSI reference model known as the “Adaptation Layer” manifested by the Overlay Multilink Network (OMNI) Interface \cite{templin2023adaptation}, which is visualized in Figure \ref{fig:templin}. The Adaptation Layer occurs logically below the Network Layer (L3) but above the Data Link Layer (L2) and employs Internet Protocol, version 6 (IPv6) as an encapsulation sublayer. The 128-bit IPv6 addressing architecture allows each node to self-select a local address that is assured to be unique within the scope of the MANET. The MANET routing protocol therefore carries these unique local addresses in its topology advertisements as fully-qualified IPv6 host routes to support the multi-hop routing and forwarding functions.

When two nodes within the same MANET wish to communicate using Internet-style applications (such as file transfer, web browsing, etc.) the OMNI interface supports direct access to the Adaptation Layer IPv6 service, and upper layer protocol framings based on the Transmission Control Protocol (TCP) or User Datagram Protocol (UDP) are carried in IPv6 packet headers with no further encapsulation applied. IPv6 packet streams between these MANET nodes can then very quickly engage AERO header compression messaging to greatly reduce precious 5G SL radio frequency bandwidth that would otherwise be consumed through transmissions of redundant IPv6 headers.

When a MANET is able to connect to a larger Internetwork through an infrastructure elements known as a “proxy”, nodes within the MANET can extend communications to peers located in distant networks (which could include both static subnetworks and other MANETs). The proxy supplies each MANET node with an Internet Protocol (IP) subnet as an unambiguous Network Layer reference point. The MANET node’s OMNI interface then employs IP-in-IPv6 encapsulation to convey packets to and from the proxy across possibly multiple MANET forwarding hops. The proxy in turn extends the transmissions across the Internetwork as a stable reference point, while the MANET topology may continue to change dynamically due to factors such as node movements. The AERO/OMNI technologies and the Adaptation Layer they introduce therefore provide key tools for effective 5G SL MANET Internetworking operations.

\section{Integrated Sensing, Computation, and Communication}
\label{se:integrated}
In many applications where direct-mode communication is of interest, it is also desirable to collect information about the environment for holistic situational awareness. For example, in vehicular applications, a vehicle may be equipped with conventional sensors such as an automotive radar, lidar, or camera, and wishes to communicate this data to a sink node for joint processing. Further, the development of full-duplex radio technology, the potential for co-located vehicular radar and communication capabilities is emerging. 

Many such vehicles will also be expected to communicate via V2X SL channels, and observations of individual users will be processed jointly. Therefore, the interplay between sensing, communication, and computational systems is a natural avenue of research and development for SL-enabled networks.

In \cite{Bartoletti2022sidelink} and \cite{Decarli2022joint}, a preliminary analysis of the integrated sensing and communication scenario is presented in the context of V2X, with a focus on fundamental performance bounds. The impact of interference and resource allocation is examined through an analysis of MCS, packet size, and vehicle density. Simulation results focus on the distribution of the Cramer-Rao Lower Bound for radar parameter estimation, and demonstrate the importance of resource allocation for both range and velocity estimation. Future studies on the integration of sensing, communication, and computation could focus on resource allocation techniques which take into account sensing or estimation-theoretic performance metrics.

In \cite{Schuhback2023sensing}, an opportunistic decentralized pedestrian sensing scheme enabled by SL communication is developed. The scheme is analyzed in the context of V2X communication and applied to the task of collective perception. It is shown that SL communication, in tandem with a beacon-based sensing mechanism, can be used to reliably reconstruct pedestrian density in a crowded vehicular environment.

\section{Machine Learning Tools and Applications}
\label{se:learning}
A potential source of improvement for multi-hop SL communication is the development of artificial intelligence and machine learning algorithms. Such learning algorithms have the potential to solve important resource allocation problems inherent to the SL scenario, particularly in SL modes 2 and 3 where coordination from a base-station is limited.

\emph{Federated Learning} (FL) is a promising methodology for distributed machine learning, which may be particularly well-suited for SL-enabled applications. Whereas conventional distributed ML models rely on sharing raw data, FL aims to preserve data privacy by sharing only the locally trained ML model parameters, such as the weights and biases of a neural network. While many early FL implementations require the use of a centralized parameter serve to orchestrate the training process, recent FL developments have exploited direct interactions between the learning agents. This decentralized learning operation is directly amenable to SL-enabled networks. In \cite{Barbieri2023layer}, a fully-decentralized, SL-compatible, FL architecture is developed. SL-enabled applications may also benefit from a semi-decentralized FL learning approach, in which a localized SL communication network is blended with the conventional device to server interaction structure.  The semi-decentralized FL framework proposed in \cite{parasnis2023conn} lends itself to SL applications involving cooperative UAV networks for intelligence, surveillance, and reconnaissance (ISR) operations in defense settings.

Another decentralized learning approach that demonstrates promise for SL resource allocation problems is the \emph{multi-player multi-armed bandit (MMAB)} framework \cite{Shi2021MMAB}. The MMAB framework models a resource allocation problem as a repeated multi-player game, and aims to facilitate coordination among the players without explicit communication, by making use of \emph{implicit communication}. Thus, the MMAB problem has proven to be useful for resource allocation problems where dedicated communication to a centralized server is impractical, such as channel allocation in cognitive radar networks \cite{howard2021multi}. In the case of tactical SL communication, the MMAB framework may allow for coordination of individual UE actions without the need for extensive relaying or reliable communication with a BS.

Future studies could focus on the feasibility of various federated and distributed learning models for resource allocation in the three modes of SL operation. Additionally, incorporation of time-sensitive metrics, such as the Age-of-information (AoI) or Age-of-incorrect-information (AoII), would be of interest for mission-critical SL applications. AoI quantifies the time elapsed since a packet of data has been generated, and is therefore an important metric when time-sensitive decisions must be made. AoII quantifies the time elapsed since an incorrect update has been generated, and is of great benefit when both the timeliness and accuracy of a measurement are important, such as the integrated sensing and communication scenario described in Section \ref{se:integrated}.

\section{Conclusion and Open Challenges}
\label{se:concl}
This work has examined several key research directions and developments related to the use of 5G SL in multi-hop tactical networking. In particular, key technical aspects of 5G SL have been summarized, potential multi-hop routing protocols have been surveyed, recent developments in open-source network emulation for SL-enabled networks has been discussed, and research directions in the areas of integrated sensing and communication and distributed learning algorithms have been proposed. 
%
%
However, the current state of 5G SL leaves several important research and development challenges open for future work including:(a) Multi-hop routing; (b) Application-specific PLA; (c) Military-grade system-level simulation; (d) Integration of communication, sensing, and computation; (e) Federated and decentralized learning algorithms for routing, resource allocation, and data processing. 

While many of the emerging research and development efforts in 5G SL demonstrate have demonstrated impressive potential, there is still a great need for many of these technology areas to be integrated into large-scale simulators and prototypes suited for test and measurement. To this end, there have been several encouraging efforts in the open-source community \cite{liu20225g,Ryu2022emane,Le2022physical,elkadi2023open,cao2023efficient}. However, to reliably meet the needs of multi-hop tactical networks, more integration efforts are required.

\bibliographystyle{IEEEtran}
\bibliography{sidelinkBib.bib}{}

\begin{thebibliography}{10}
\providecommand{\url}[1]{#1}
\csname url@samestyle\endcsname
\providecommand{\newblock}{\relax}
\providecommand{\bibinfo}[2]{#2}
\providecommand{\BIBentrySTDinterwordspacing}{\spaceskip=0pt\relax}
\providecommand{\BIBentryALTinterwordstretchfactor}{4}
\providecommand{\BIBentryALTinterwordspacing}{\spaceskip=\fontdimen2\font plus
\BIBentryALTinterwordstretchfactor\fontdimen3\font minus
  \fontdimen4\font\relax}
\providecommand{\BIBforeignlanguage}[2]{{%
\expandafter\ifx\csname l@#1\endcsname\relax
\typeout{** WARNING: IEEEtran.bst: No hyphenation pattern has been}%
\typeout{** loaded for the language `#1'. Using the pattern for}%
\typeout{** the default language instead.}%
\else
\language=\csname l@#1\endcsname
\fi
#2}}
\providecommand{\BIBdecl}{\relax}
\BIBdecl

\bibitem{Rettore2022towards}
P.~Rettore \emph{et~al.}, ``Towards software-defined tactical networks:
  Experiments and challenges for control overhead,'' in \emph{IEEE Mil. Commun.
  Conf.}, 2022, pp. 110--116.

\bibitem{Burbank2006}
J.~L. Burbank \emph{et~al.}, ``Key challenges of military tactical networking
  and the elusive promise of {MANET} technology,'' \emph{IEEE Commun. Mag.},
  vol.~44, no.~11, pp. 39--45, 2006.

\bibitem{elmasry2021operational}
G.~Elmasry and P.~Corwin, ``Operational views of vertical tactical {5G},'' in
  \emph{IEEE Mil. Commun. Conf.}, 2021, pp. 727--732.

\bibitem{bastos2021potential}
L.~Bastos \emph{et~al.}, ``Potential of {5G} technologies for military
  application,'' in \emph{Intl. Conf. Mil. Commun. Inform. Syst.}\hskip 1em
  plus 0.5em minus 0.4em\relax IEEE, 2021, pp. 1--8.

\bibitem{campolo2019impact}
C.~Campolo \emph{et~al.}, ``{5G NR V2X}: On the impact of a flexible numerology
  on the autonomous sidelink mode,'' in \emph{IEEE 5G World Forum}, 2019, pp.
  102--107.

\bibitem{Bartoletti2022sidelink}
S.~Bartoletti, N.~Decarli, and B.~M. Masini, ``Sidelink {5G-V2X} for integrated
  sensing and communication: the impact of resource allocation,'' in \emph{IEEE
  ICC Workshops}, 2022, pp. 79--84.

\bibitem{Draper2011cooperative}
S.~C. Draper, L.~Liu, A.~F. Molisch, and J.~S. Yedidia, ``Cooperative
  transmission for wireless networks using mutual-information accumulation,''
  \emph{IEEE Trans. Inf. Thy.}, vol.~57, no.~8, pp. 5151--5162, 2011.

\bibitem{liu20225g}
P.~Liu \emph{et~al.}, ``{5G} new radio sidelink link-level simulator and
  performance analysis,'' in \emph{ACM Conf. on Modeling Analysis and Simul. of
  Wirel. Mobile Syst.}, 2022, pp. 75--84.

\bibitem{Ryu2022emane}
B.~Ryu \emph{et~al.}, ``{5G-EMANE}: Scalable open-source real-time {5G} new
  radio network emulator with {EMANE},'' in \emph{IEEE Mil. Commun. Conf.},
  2022, pp. 553--558.

\bibitem{Le2022physical}
A.~D. Le \emph{et~al.}, ``Physical layer abstraction for {LTE} and {5G} new
  radio with imbalance receiver in {EMANE},'' in \emph{IEEE Mil. Commun.
  Conf.}, 2022, pp. 515--521.

\bibitem{elkadi2023open}
M.~Elkadi \emph{et~al.}, ``Open source-based over-the-air {5G} new radio
  sidelink testbed,'' \emph{arXiv preprint arXiv:2306.09286}, 2023.

\bibitem{cao2023efficient}
L.~Cao, L.~Zhang, and S.~Roy, ``Efficient {PHY} layer abstraction for {5G NR}
  sidelink in ns-3,'' in \emph{Proc. 2023 Workshop on ns-3}, 2023, pp.
  115--120.

\bibitem{templin2023adaptation}
F.~L. Templin, A.~Agarwal, M.~M. Badgandi, and B.~R.~S. Prakash, ``An
  adaptation layer for the internet: The “6 m's of modern
  internetworking”,'' in \emph{2023 IEEE Aerospace Conference}.\hskip 1em
  plus 0.5em minus 0.4em\relax IEEE, 2023, pp. 1--10.

\bibitem{Decarli2022joint}
N.~Decarli, S.~Bartoletti, and B.~M. Masini, ``Joint communication and sensing
  in {5G-V2X} vehicular networks,'' in \emph{IEEE Medit. Electrotechnical
  Conf.}, 2022, pp. 295--300.

\bibitem{Schuhback2023sensing}
S.~Schuhbäck, L.~Wischhof, and J.~Ott, ``Cellular sidelink enabled
  decentralized pedestrian sensing,'' \emph{IEEE Access}, vol.~11, pp.
  13\,349--13\,369, 2023.

\bibitem{Barbieri2023layer}
L.~Barbieri, S.~Savazzi, and M.~Nicoli, ``A layer selection optimizer for
  communication-efficient decentralized federated deep learning,'' \emph{IEEE
  Access}, vol.~11, pp. 22\,155--22\,173, 2023.

\bibitem{parasnis2023conn}
R.~Parasnis \emph{et~al.}, ``Connectivity-aware semi-decentralized federated
  learning over time-varying {D2D} networks,'' in \emph{Intl. Conf. on Mobile
  Comput. Netw.}\hskip 1em plus 0.5em minus 0.4em\relax ACM, 2023, pp. 1--10.

\bibitem{Shi2021MMAB}
C.~Shi and C.~Shen, ``Multi-player multi-armed bandits with collision-dependent
  reward distributions,'' \emph{IEEE Trans. Signal Process.}, vol.~69, pp.
  4385--4402, 2021.

\bibitem{howard2021multi}
W.~W. Howard \emph{et~al.}, ``Multi-player bandits for distributed cognitive
  radar,'' in \emph{IEEE Radar Conf.}\hskip 1em plus 0.5em minus 0.4em\relax
  IEEE, 2021, pp. 1--6.

\end{thebibliography}

\end{document}